 \documentclass[final,5p,times,twocolumn]{elsarticle}

\usepackage{graphicx}
\usepackage{amssymb}
\usepackage{amsmath}

\usepackage{hyperref}
\hypersetup{
    unicode=false,          % non-Latin characters in Acrobat’s bookmarks
    pdftoolbar=true,        % show Acrobat’s toolbar?
    pdfmenubar=true,        % show Acrobat’s menu?
    pdffitwindow=false,     % window fit to page when opened
    pdfstartview={FitH},    % fits the width of the page to the window
    pdftitle={Perspectives of optical lattices with state-dependent tunneling},    % title
    pdfauthor={Andrii Sotnikov},     % author
    colorlinks=true,        % false: boxed links; true: colored links
    linkcolor=red,          % color of internal links (change box color with linkbordercolor)
    citecolor=green,        % color of links to bibliography
    filecolor=magenta,      % color of file links
    urlcolor=cyan           % color of external links
}

\usepackage{color} 				%% colored text
\usepackage[normalem]{ulem} 			%% crossed text
\newcommand{\as}[1]{#1}
\newcommand{\asc}[2]{{}{#2}}

\begin{document}

\begin{frontmatter}



%\title{Perspectives of state-dependent optical lattices in approaching quantum magnetism in the presence of the external harmonic trapping potential}
\title{Perspectives of optical lattices with state-dependent tunneling in approaching quantum magnetism in the presence of the external harmonic trapping potential}
\author{Andrii Sotnikov}
\ead{a\_sotnikov@kipt.kharkov.ua}
\address{Akhiezer Institute for Theoretical Physics, NSC KIPT, Akademichna Str. 1, 61108 Kharkiv, Ukraine}
\address{Karazin Kharkiv National University, Svobody Sq. 4, 61022 Kharkiv, Ukraine}
%\affiliation{Institut f\"ur Theoretische Physik, Goethe-Universit\"at, 60438 Frankfurt/Main, Germany}


%% ABSTRACT %%
\begin{abstract}
%Motivated by a successful realization of state-dependent optical lattices for two-component mixtures of ultracold potassium-40 atoms with the magnetic-field-gradient technique, 
We study theoretically potential advantages of two-component mixtures in \as{optical lattices with state-dependent tunneling} for approaching long-range-order phases and detecting easy-axis antiferromagnetic correlations. 
%% 37
%Our analysis is based on the dynamical mean-field theory with the local-density approximation that accounts for inhomogeneity effects produced by the external trapping potential.
%By comparing two opposite limits in the experimental realizations, we conclude that, while the strongly-imbalanced mixtures are advantageous for approaching quantum magnetism in homogeneous systems, 
%We find that in the harmonic trap the effects originating from the entropy redistribution become highly important. magnetic states require lower entropy per particle, thus lower initial temperatures due to smaller amount of entropy that can be distributed in metallic shells surrounding the bulk. 
While we do not find additional advantages of mixtures with large hopping imbalance for approaching quantum magnetism in a harmonic trap, 
it is shown that a nonzero difference in hopping amplitudes remains highly important for a proper symmetry breaking in the pseudospin space for the single-site-resolution imaging and can be advantageously used for a significant increase of the signal-to-noise ratio and thus 
detecting long-range easy-axis antiferromagnetic correlations in the corresponding experiments.
\end{abstract}


\end{frontmatter}

%\pacs{67.85.-d, 37.10.Jk, 71.10.Fd, 75.10.Jm, }%, 71.45.Lr -- CDW systems, 75.50.Ee -- antiferromagnetism; 75.10.Jm --
%\maketitle

%% INTRODUCTION %%
\section{Introduction}
Due to a recent experimental realization of state-dependent optical lattices for two-component mixtures of ultracold $^{40}$K atoms with the magnetic-field-gradient technique and low heating rate \cite{Jotzu2015PRL} it is now much easier to access and study asymmetric lattice models without a necessity of using heteronuclear fermionic mixtures (e.g., $^{6}$Li--$^{40}$K) or long-living metastable electronic states of the same fermionic isotope (e.g., $3P_0$ state of $^{173}$Yb).
Among potential applications of this technique one can suggest approaching long-range magnetically-ordered states \cite{Sotnikov2012PRL}. It is known that quantum magnetism in ultracold fermionic mixtures is one of major experimental challenges nowadays and a significant progress already has been made in this direction. In particular, short-range antiferromagnetic (AFM) correlations were effectively measured \cite{Greif2013S,Har2015Nat} and their unique dynamics in the presence of the tunable lattice geometry was observed recently \cite{Greif2015PRL}. 

Considering two-component fermionic mixtures from the point of view of theoretical models and spin symmetries, \as{optical lattices with state-dependent (i.e., spin-dependent) hopping amplitudes} effectively break the initial continuous SU(2) symmetry of the system described by the Hubbard model towards U(1)$\times\mathbb{Z}_2$, where $\mathbb{Z}_2$ is a discrete reflection symmetry along the easy axis. The easy-axis direction is important, in particular, for the experimental detection of AFM correlations based \as{on the Bragg spectroscopy analysis \cite{Har2015Nat,Corcovilos2010PRA} and} on the quantum-gas-microscope technique (QGMT).
Despite the fact that temperatures\as{ and entropies}\footnote{\as{The entropy per particle is more crucial quantity in the context of ultracold-atom experiments, since the system does not exchange heat with environment and its parameters can be changed adiabatically.}} achieved with recent successful developments of the QGMT for fermionic mixtures \cite{Haller2015,Cheuk2015,Parsons2015,Edge2015PRA,Orman2015PRL,Greif2015pre2}
are high to observe long-range magnetic correlations, thus further optimizations and improvements in cooling protocols are required, it is important to study characteristic dependencies of\as{ these} thermodynamic quantities\asc{ such as temperature and entropy per particle}{} on other system parameters (including different symmetries of magnetic ground states), thus determine the most optimal regime for {\it in situ} imaging of the long-range AFM correlations.

%% SYSTEM AND MODEL %%
\section{Theoretical description}\label{sec.2}
Ultracold two-component fermionic mixtures in optical lattices with a sufficiently strong lattice potential, $V_{\text{lat}}\gtrsim 5E_{\text{r}}$, where $E_{\text{r}}$ is the recoil energy of atoms, are well described by the single-band Hubbard model with the Hamiltonian
\begin{eqnarray}\label{H}
\mathcal{\hat{H}}=&&
-\sum\limits_{\langle ij\rangle}\sum\limits_{\sigma} t_\sigma (\hat{c}^\dag_{i\sigma}\hat{c}_{j\sigma}+{\rm h.c.})
+U\sum\limits_{i}\hat{n}_{i\uparrow}\hat{n}_{i\downarrow}
\nonumber\\
&&+V\sum\limits_{i}(r_i/a)^2\hat{n}_{i}
-\as{\sum\limits_{i}\sum\limits_{\sigma}\mu_\sigma\hat{n}_{i\sigma} },
\label{eq.1}
\end{eqnarray}
where $t_\sigma$ is the hopping amplitude of fermionic species \as{in a particular atomic hyperfine state that we denote by the spin-1/2 index $\sigma=\{\uparrow,\downarrow\}$, thus we consider here and below two atomic components as pseudospins},
$\hat{c}^\dag_{i\sigma}$ ($\hat{c}_{i\sigma}$) is the corresponding creation (annihilation) operator of atoms at the lattice site~$i$,
the notation $\langle ij\rangle$ indicates a summation over nearest-neighbor sites, and $U$ is the magnitude of the on-site repulsive ($U>0$) interaction of the two different species with corresponding densities $\hat{n}_{i\uparrow}$ and $\hat{n}_{i\downarrow}$  ($\hat{n}_{i\sigma}=\hat{c}^\dag_{i\sigma}\hat{c}_{i\sigma}$). 
In the third term, $V$ is the amplitude of the external harmonic potential, $r_i$ is the distance from the lattice site $i$ to the trap center, and $a$ is the lattice spacing. In the last term, \as{$\mu_\sigma$} is the chemical potential that determines the total number of atoms\as{ of each spin component} in the system. 

\as{
Note that the asymmetric Hubbard model~(\ref{H}) (it can also be recognized as the extended Falicov--Kimball model for spinless fermions in the context of solid-state materials \cite{Batista2002PRL}) can be transformed to an anisotropic Heisenberg (or a spin-1/2 XXZ) model in the limit of $U/t_\sigma\gg1$ and $n_i\approx1$. The latter is described by the Hamiltonian
\begin{eqnarray}\label{Heff}
 \mathcal{\hat{H}}_{XXZ}&=&J_{\parallel}\sum_{\langle ij\rangle}\hat{S}_{i}^{Z}\hat{S}_{j}^{Z}
 +J_{\perp}\sum_{\langle ij\rangle}(\hat{S}_{i}^{X}\hat{S}_{j}^{X}+\hat{S}_{i}^{Y}\hat{S}_{j}^{Y})
 \nonumber
\\
 && - \Delta\mu\sum_i \hat{S}_{i}^{Z},
\end{eqnarray}
with the constants $J_{\parallel}=2(t_\uparrow^2+t_\downarrow^2)/U$, $J_{\perp}=4t_\uparrow t_\downarrow/U$, $\Delta\mu=(\mu_\uparrow-\mu_\downarrow)$, and the spin-1/2 operators $\hat{S}^R_{i}= \frac{1}{2}\hat{c}^\dag_{i\alpha}\sigma^R_{\alpha\beta}\hat{c}_{i\beta}$ (here and below we use units $\hbar=1$), where $\sigma^R$ are the Pauli matrices ($R=\{X,Y,Z\}$).
Hence, we see that the presence of hopping imbalance ($t_\uparrow\neq t_\downarrow$) results in $J_{\parallel}>J_{\perp}$, thus breaks the SU(2) rotational spin symmetry towards U(1)$\times\mathbb{Z}_2$, where the discrete symmetry $\mathbb{Z}_2$ can be broken either spontaneously by long-range AFM ordering along the $Z$ axis or by the chemical potential difference $\Delta\mu\neq0$ that plays a role of the external magnetic field favoring the ferromagnetic configuration along the same axis. Naturally, with an increase of $\Delta\mu$ at the fixed asymmetry in hopping amplitudes the easy-axis (Ising-type) AFM configuration becomes less and less energetically favorable, thus a transition to another AFM-ordered easy-plane (canted) many-body state becomes possible (see Refs.~\cite{Gottwald2009PRA,Wunsch2010PRA,Sotnikov2013PRA} for more details). 

Therefore, to avoid a potential competition between different types of AFM ordering that can also result in a significant suppression of critical temperatures, in the original model~(\ref{H}) we consider chemical potentials} the same for both spin components (i.e., $\mu_\uparrow=\mu_\downarrow\equiv\mu$). 
%Note that results in $N_\uparrow=N_\downarrow$ for the balanced mixture ($t_\uparrow=t_\downarrow$). 
Note that asymmetric hopping amplitudes $t_\uparrow\neq t_\downarrow$ together with $\mu_\uparrow=\mu_\downarrow$ result in\as{ a nonzero polarization, i.e. not equal total numbers of particles in two spin states ($N_\uparrow\neq N_\downarrow$), of the trapped system \cite{Sotnikov2012PRL,Sotnikov2013PRA},} %in general case, 
but this condition is the most optimal for the easy-axis AFM ground state of the model~(\ref{H}) at half filling (e.g., at $\mu=U/2$ and $r_i=0$)\as{, as discussed above}.

Below, we consider a three-dimensional optical lattice setup with the Hubbard parameters entering Eq.~(\ref{H}) that are set close to the experimental values~\cite{Jotzu2015PRL}. In particular, we focus on two opposite limits: (i) zero (or very small) hopping imbalance $t_\uparrow= t_\downarrow=t$ and (ii) large hopping imbalance (e.g., $t_\uparrow=0.54t$ and $t_\downarrow=0.06t$). According to Ref.~\cite{Jotzu2015PRL}, both cases can be effectively realized by the magnetic-field-gradient technique with a high level of control.

%% METHOD %%
%\section{Method}\label{sec.3}
Our theoretical analysis is based on the dynamical mean-field theory (DMFT) \cite{Geo1996RMP} with the exact diagonalization solver \cite{Caffarel1994PRL}\as{ and the number of orbitals $n_s=5$ per each spin component in the corresponding Anderson impurity model. Since we are interested mostly in the easy-axis observables, in DMFT it is enough to account for the standard hybridization terms between the impurity and the bath \cite{Caffarel1994PRL}. The corresponding Anderson parameters of the impurity model are found iteratively till the convergence based on DMFT self-consistency conditions \cite{Geo1996RMP} is reached}. For the given values of the Hubbard parameters \as{our approach} allows to calculate the local observables, such as expectation values of the density of particles of any spin component, the double and the hole occupancy, as well as the fluctuations of the particle number on a particular lattice site. 

To account for the inhomogeneity effects produced by the external trap, we use DMFT with the local density approximation (LDA+DMFT). Note that LDA does not account for the proximity effects close to the phase boundaries, however in the cases under study these effects do not play a crucial role leading only to minor corrections. 
Within LDA we obtain the local observables at the specific point $r$ of the trap from the condition
$
 \mu(r)=\mu_0 - V(r/a)^2$, where $\mu_0$ is the chemical potential in the trap center that for the fixed values of the Hubbard parameters defines also the total number of particles in the system.

From the converged solutions of LDA+DMFT on different lattice sites (i.e., with different $r$) one can analyze the dependence of the local observables on the distance $r$ \as{(see also Ref.~\cite{Ho2010Nat} for more details)}. In particular, by combining the results with the Maxwell relation $\partial s/\partial \mu = \partial n/\partial T$ we obtain the entropy per lattice site at the particular point $r_0$ of the trap (for simplicity, we use the units of $k_B=1$ and $a=1$ below)
\begin{equation}\label{entr}
 s(r_0)=2V\int_{r_0}^{R_{\max}}\dfrac{\partial n(r,U,T)}{\partial T}rdr,
\end{equation}
where the cut-off distance $R_{\max}$ is determined from the condition $n(R_{\max},U,T)=0$.

A subsequent integration of the entropy and density distributions in the trap determine the total entropy $S$ and the total number of particles $N$ in the system (here and below we assume the axial-symmetric three-dimensional setup)
\begin{equation}\label{entr_tot}
 S = \int_{0}^{R_{\max}}s(r)4\pi r^2dr,\quad N = \int_{0}^{R_{\max}}n(r)4\pi r^2dr.
\end{equation}
Both quantities, $S$ and $N$, can be considered as the preserved numbers in the experiment (and, in particular, during the lattice ramp) that allows to access the initial values for the entropy and temperature. Note that, alternatively, one can also introduce an additional term corresponding to the amount of entropy per particle $\Delta s$ that is added to the system due to uncontrolled heating processes during the lattice ramp, as it was done in Ref.~\cite{Greif2013S}. Below, for simplicity and consistency reasons, we consider that the change in system parameters can be performed adiabatically ($\Delta s=0$).

In order to determine the initial temperature $\widetilde{T}$ in the system that is necessary for observations of the many-body quantum phases under study we use the expression for the entropy of the Fermi gas under assumption of a moderate scattering length $a_s$ ($k_{\text{F}} |a_s|<1/2$) \cite{Carr2004PRL}
\begin{equation}\label{entr_Carr}
 S\approx N \pi^2 \widetilde{T}/{T_{\text{F}}}.
\end{equation}
Eqs.~(\ref{entr})-(\ref{entr_Carr}) allow to set a direct correspondence between thermodynamic quantities before and after the lattice ramp. Therefore, the problem can be effectively solved under assumption of adiabaticity of the ramp process.

\section{Results}\label{sec.4}

\subsection{Comparative entropy analysis of the trapped mixtures including hopping imbalance}
To proceed with the quantitative analysis, let us specify the Hubbard parameters, which we set maximally close to the experimental values~\cite{Jotzu2015PRL}. In particular, to simplify the theoretical description and in accordance with Eqs.~(\ref{entr})-(\ref{entr_Carr}), we consider the axial-symmetric three-dimensional trap with the frequency $\bar{\omega}= 2\pi\times68.4$~Hz (i.e., the geometric mean of three frequencies $\omega_{1,2,3}$ in \cite{Jotzu2015PRL}). By specifying $t=174$~Hz \cite{Jotzu2015PRL}, we obtain the amplitude of the trapping potential $V=0.015t$ in Eq.~(\ref{H}). Concerning imbalances in hopping amplitudes, we focus below on two realizations: (i) $t_\uparrow=t_\downarrow=t$ and (ii) $t_\uparrow=0.54t$ and $t_\downarrow=0.06t$. (Since $V$ is scaled in units of $t$, the trap confinement is effectively stronger in the second realization.) Next, we assume that the number of particles of each spin component $N_\sigma$ can be tuned independently, as well as the interaction strength $U$ 
between components can be controlled separately by means of Fesh\-bach resonances. In other words, one can always choose the optimal Hubbard parameters $\mu$ and $U$ from the conditions to approach and observe the largest AFM-ordered domains at a given initial temperature. According to the studies of two-component fermionic mixtures with hopping imbalance~\cite{Sotnikov2012PRL} we can set $\mu_0=U/2$, thus expect these domains in the trap center, and $U=10t^*$ with $t^*=(t_\uparrow+t_\downarrow)/2$ that leads to appearance of AFM correlations at high enough temperatures (up to $T\approx0.5t^*$) in both cases. 

\begin{figure}
\includegraphics[width=\linewidth]{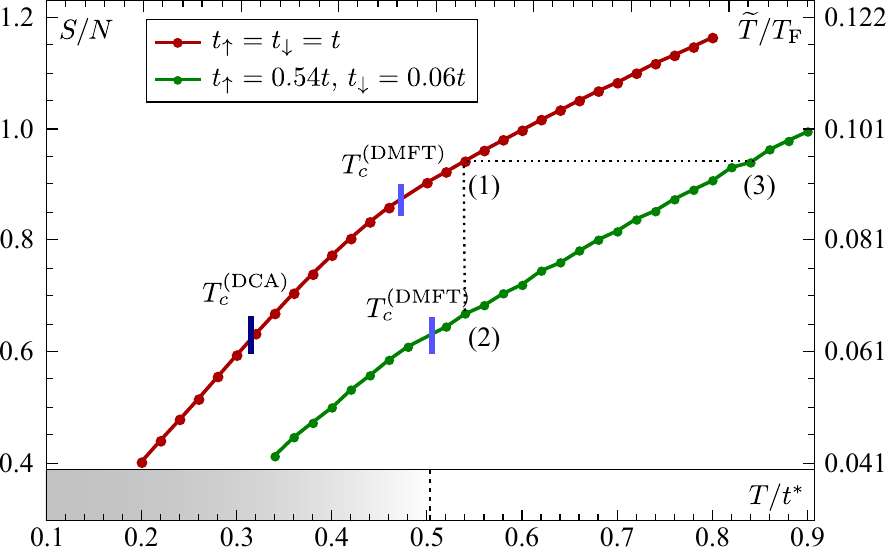}
\caption{Dependence of the entropy per particle (left axis) and the initial temperature (under assumption of $\Delta s\rightarrow0$; right axis) on the temperature in a simple cubic lattice obtained by DMFT for two-component mixtures with two different realizations of imbalance in hopping amplitudes\as{, $U=10t^*$, and $V=0.015t$}. \as{The gray-shaded bar in the bottom part qualitatively indicates the strength of the AFM correlations in the regions with $n\approx1$.} The reference value of \as{$T_c^{\text{(DCA)}}$} for the balanced mixture is taken from Ref.~\cite{Kent2005PRB}.}\label{fig1}
\end{figure}

\begin{figure}
\includegraphics[width=\linewidth]{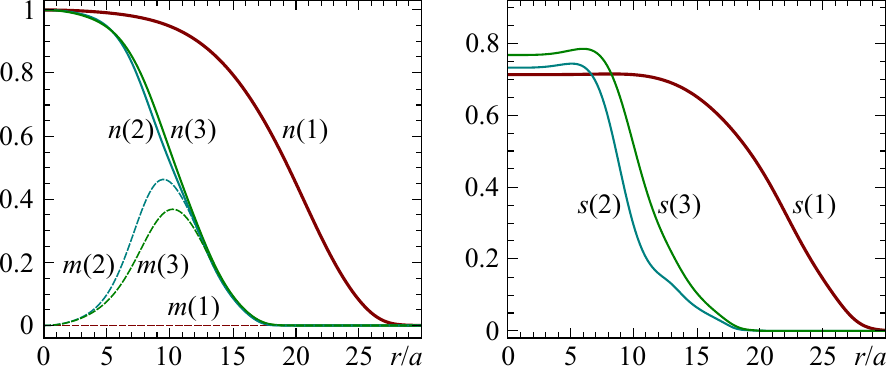}
\caption{Real-space distributions of the on-site filling $n$, magnetization $m\equiv({n}_{i\uparrow}-{n}_{i\downarrow})$, and the entropy per lattice site $s$ in the harmonic trap with $V=0.015t$ obtained by LDA+DMFT in three specific points chosen from Fig.~\ref{fig1}.}\label{fig2}
\end{figure}

In Fig.~\ref{fig1}
we show the dependence of the entropy per particle $S/N$ and, therefore, the initial temperature $\widetilde{T}$ of the trapped Fermi gas before the lattice ramp on the temperature in the lattice $T$. We conclude that in mixtures with large hopping imbalance (the second realization) it is necessary to start with the lower initial temperature to approach the regime with AFM correlations. This may seem contradictory to expectations based on the entropy analysis of homogeneous systems~\cite{Sotnikov2012PRL}. However, the effect becomes clear from the analysis of real-space distributions that are shown in Fig.~\ref{fig2}.
As one can notice, the effect originates from a smaller amount of entropy that can be distributed in mixtures with the hopping imbalance due to significant magnetization of metallic shells in the presence of the trapping potential. 
Note that, according to an additional analysis performed and due to the real-space entropy distributions shown in Fig.~\ref{fig2}, reducing the magnitude of the trapping potential in $t/t^*$ times and a corresponding increase of the total number of particles in the system does not help to suppress the main effect.

Concerning a potential compensation of ferromagnetic shells, their presence is required to guarantee a proper parameter regime in the bulk that is optimal for the easy-axis AFM state. Thus adding more particles of the less mobile (spin-down) component to the system will result not only in a competition between different types of AFM ordering in the bulk, but also in two ferromagnetic shells that are usually separated in real space (see Ref.~\cite{Sotnikov2013PRA} for details). The latter fact does not allow to significantly increase the entropy $S/N$ necessary for observations of long-range AFM correlations.

Therefore, inducing strong asymmetry in hopping amplitudes of two-component fermionic mixtures in a trap does not result in additional advantages from the point of view of approaching the AFM-ordered regime with the highest possible initial temperature before the lattice ramp. However, a small (nonzero) asymmetry in hopping amplitudes may be still a very important ingredient for observations of the long-range AFM correlations as it is shown below.

\subsection{Direct observations of AFM correlations: importance of a specific direction for the symmetry breaking}
Let us consider the realization when the hopping amplitudes of both spin components can be considered as (a) approximately or (b,c) strictly equal each other. In the first case, we consider that the difference in the hopping amplitudes is small not to affect thermodynamic properties of the mixture, so that one can follow the obtained dependence for the balanced mixture shown in Fig.~\ref{fig1}, but it is large enough to break the symmetry in the Hamiltonian~(\ref{H}) in a controlled way. According to Refs.~\cite{Sotnikov2012PRL,Sotnikov2013PRA}, at low temperatures the system prefers the easy-axis AFM-ordered configuration. In another limit of strictly equal hopping amplitudes we consider two realizations: (b) a small population imbalance is present, so that the system prefers the easy-plane AFM configuration, and (c) the mixture is balanced completely, so that the AFM state corresponds to spontaneous breaking of the SU(2) rotational symmetry in the pseudospin space.

In the analysis we restrict ourselves to measurements based on QGMT, thus spatial correlations involving number of particles of each spin component on all lattice sites that belong to a specific plane of the three-dimensional setup can be directly observed. It means that the easy-axis correlations (in the $Z$ direction of the three-dimensional pseudospin space) can be directly measured, but to obtain some signal from the easy-plane (XY-AFM) correlations one needs, e.g., an additional $\pi/2$-rotation in the pseudospin space to be performed beforehand \as{(see also Ref.~\cite{Corcovilos2010PRA} for details)}. The latter in ultracold mixtures of alkali-metal atoms is usually performed by applying additional rf-pulses. 
Therefore, from the experimental point of view, the above realizations of the symmetries correspond to 
(a) controlled both small nonzero hopping imbalance and the corresponding population difference (to produce weak ferromagnetic metallic shells in a trap), 
(b) zero (up to the experimental accuracy) hopping and small nonzero population imbalance, 
and (c) balanced mixture in both respects.

\as{Now, to proceed with quantitative estimates in the framework of DMFT, let us specify the realizations (a) and (b). 
For the case (a) we take $t_\uparrow=1.05 t$, $t_\downarrow=0.95t$, and $\mu_\uparrow=\mu_\downarrow$. With the amplitude $V=0.015t$ this results in the polarization $P=(N_\uparrow-N_\downarrow)/(N_\uparrow+N_\downarrow)\approx0.09$ that remains almost constant with the temperature $T/t$ (or, analogously, $S/N$) change in the region of interest.
Hence, for the case (b) we take $t_\uparrow=t_\downarrow=t$ and $P\approx0.09$ that corresponds to $\Delta\mu=0.07t$. 

%According to an additional analysis (the method is introduced in Ref.~\cite{Sotnikov2013PRA}) we found that the specified asymmetries unambiguously set the proper direction for AFM ordering, as discussed above, but insignificantly affect main thermodynamic characteristics of the system. Moreover, at the same entropy per particle $S/N$ (more precisely, within a deviation of less than 1\% for $S/N$ value) the dependence of the main quantity of interest, the on-site magnetization ${\bf m}_i\equiv(S_i^X,S_i^Y,S_i^Z)$,

Next, by having access to the main quantity of interest, the on-site magnetization ${\bf m}_i\equiv(\langle \hat{S}_i^X\rangle,\langle \hat{S}_i^Y\rangle,\langle \hat{S}_i^Z\rangle)$ as a function of the distance~$r$ to the trap center, we can reproduce the expected single-site-resolution images for the occupancy by the single (e.g., spin-up) atomic component at specific values of $S/N$.
To this end, we perform a Monte-Carlo (MC) sampling, where $(m^Z_i+n_i/2)$ is used as a weighting factor for occupation of the lattice site~$i$ by the spin-up particle. In particular, by considering two limiting cases at $n=1$, (i) $m^Z_i=\pm1/2$, the site is always occupied (unoccupied) by the spin-up component, thus the checkerboard pattern is reproduced on the site~$i$; (ii) $m^Z_i=0$, the site~$i$ resembles the checkerboard structure or not with equal probabilities.

Note that, while from the point of view of computational analysis and experimental observations the images for the case (a) can be obtained now in a straightforward way due to the easy-axis AFM order, in the case (b) the ordering takes place in the $XY$ plane. Therefore, in the experiment one needs to perform an additional $\pi/2$ rotation in the pseudospin space. At the same time, in the computational procedure, where the rotational $U(1)$ symmetry in the $XY$ plane is broken to simplify the analysis (e.g., in the $X_0$ direction), one needs to account for uncontrolled degrees of freedom (rotational symmetry) by multiplying the magnetization by a numerical factor that corresponds to the averaging of the projection length of a random vector on the $X_0$ axis, thus  $m^Z_i= 2m^{X_0}_i/\pi $.
In the case (c) no additional rotations are necessary from the experimental point of view, but from the computational side the same arguments must be applied for the SU(2)-symmetric system, thus $m^Z_i= m^{Z_0}_i/2 $, where $Z_0$ denotes the fixed direction in the numerical analysis.
}

\begin{figure}
\includegraphics[width=\linewidth]{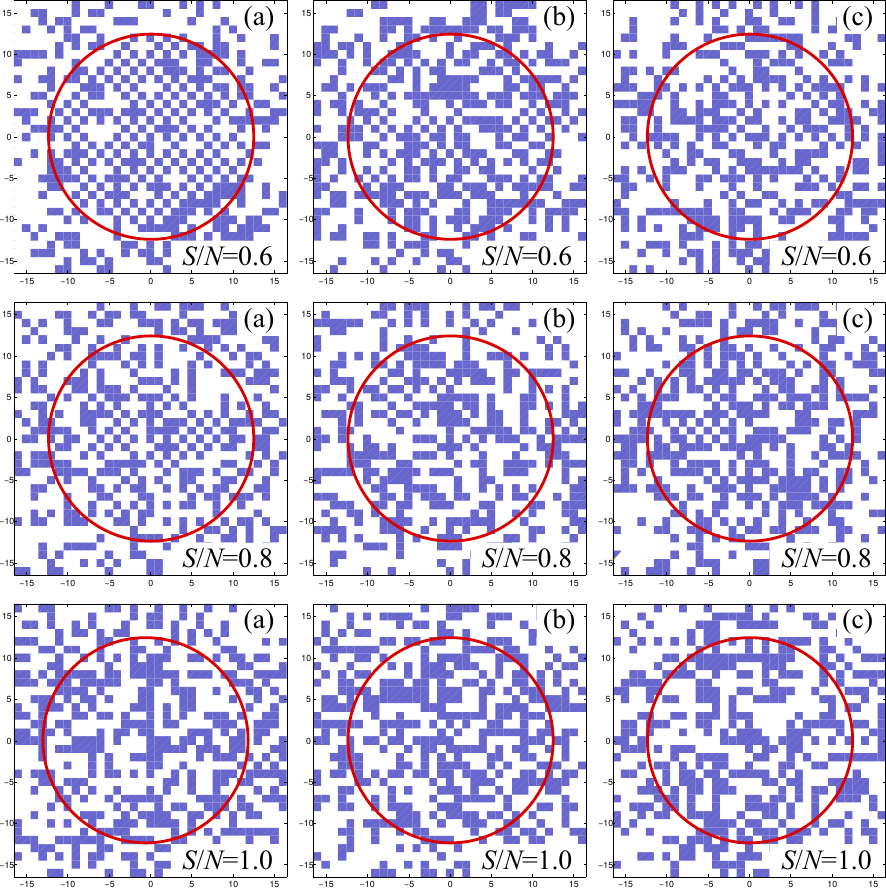}
\caption{Characteristic real-space single-shot images of the in-plane occupation of lattice sites by one spin component at different entropies per particle and different experimental realizations (different types of the controlled symmetry breaking) in a trap. The regions with $n\approx1$, where AFM correlations can be present, are surrounded by circles. The images are taken close to the most probable values of $F/F_0$ (see Fig.~\ref{fig4}) from the corresponding sets of MC samples.}\label{fig3}
\end{figure}

\as{By considering all three cases (a)-(c) in} Fig.~\ref{fig3}
we directly see, that the effect produced by the controlled symmetry breaking can be significant. In particular, we conclude that in the realization~(a) at $S/N=0.6$ the long-range correlations are much easier to observe directly by means of QGMT \as{or Bragg scattering}.

\begin{figure}
\includegraphics[width=\linewidth]{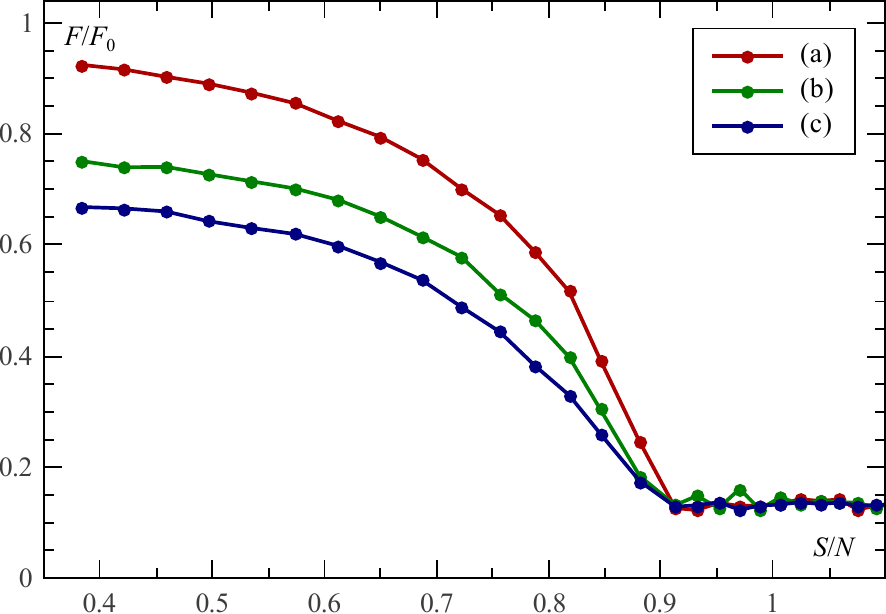}
\caption{Dependencies of the mean size of the largest in-plane AFM domain inside the region $r\leq12a$ (with the total number of sites $F_0= 441$) on the entropy per particle in the system. Each point corresponds to averaging over 100 MC samples (real-space images).}\label{fig4}
\end{figure}

Finally, assuming that in the experiment one can perform an additional statistical analysis, in other words, one can average over some finite set of real-space images (similar to those shown in Fig.~\ref{fig3}) with the fixed system parameters (including temperature and the realization type), we can determine the average size of the largest domain and analyze its dependence on the temperature and the type of the controlled symmetry breaking.
For this purpose, we apply a procedure similar to the Hoshen--Kopelman algorithm in percolation theory \cite{Hoshen1976PRB} with accounting for the checkerboard structure of domains and summarize our results in Fig.~\ref{fig4}.

We conclude that this statistical analysis helps to extract more useful information about the AFM correlations in addition to the single-shot images shown in Fig.~\ref{fig3}. In all three realizations (a)-(c) the effects produced by magnetic correlations can be well identified and separated from the uncorrelated background signal (with $F/F_0\approx0.13$ in Fig.~\ref{fig4}). However, even in this case we see a significant advantage of the realization (a) that has the largest signal-to-noise ratio in the whole region of entropies below $S/N\approx0.9$\footnote{\as{Additional estimates based on DMFT analysis show that for a two-dimensional system with square lattice geometry the obtained dependencies remain qualitatively the same with a corresponding decrease in values of $S/N$ (depending on the parameters of a particular experimental setup) approximately by 20\%.}}.

\section{Conclusions}
We studied theoretically perspectives of state-dependent optical lattices realized in Ref.~\cite{Jotzu2015PRL} for the purpose of approaching and observing many-body quantum states with long-range magnetic correlations in ultracold two-component fermionic mixtures. We conclude that with account of the external harmonic potential there are no additional advantages of mixtures with large imbalance in hopping amplitudes for approaching quantum magnetism at the fixed value of the entropy per particle in the system. The strongest effect in this case originates from the structure of metallic shells that have nonzero magnetization and thus contain less entropy than in the balanced case.

We analyzed a possibility to control the symmetry-breaking direction by using state-dependent optical lattices. It is shown that systems with small nonzero hopping imbalance have clear advantages among others for observations based on measurements of the particle number and, in particular, in the developed experimental techniques involving quantum gas microscope for ultracold fermions \as{\cite{Haller2015,Cheuk2015,Parsons2015,Edge2015PRA,Orman2015PRL,Greif2015pre2}}. We obtained characteristic single-shot images and performed statistical analysis of their structure that allowed us to determine characteristic values for the entropy per particle corresponding to appearance of sizable AFM-ordered domains in the system.

\as{The dependence on the smallness of asymmetries required for a controlled symmetry breaking can influence nonequilibrium properties such as thermalization time for approaching a particular magnetically-ordered state. In this paper, by focusing only on equilibrium characteristics, we assumed that the introduced magnitudes of asymmetries in hopping amplitudes and number of particles are large enough for the system to equilibrate within the experimental timescale. Therefore, nonequilibrium properties from that perspective remain highly important and could be studied in detail with further extensions \cite{Tsuji2013PRL} of the used theoretical approach.
}

%\begin{acknowledgments}
\section*{Acknowledgments}
The author thanks Walter Hofstetter for suggesting the research direction and Gregor Jotzu for fruitful discussions and providing information about their experimental possibilities. This work is partly supported by the National Fund of Fundamental Research of Ukraine, grant No. 25.2/102.
%\end{acknowledgments}

%

\end{document}